\documentclass[%
 reprint,
 superscriptaddress,
 amsmath,amssymb,
 aps,
]{revtex4-1}

\usepackage{threeparttable}
\usepackage{braket}
\usepackage{graphicx}
\usepackage{dcolumn}
\usepackage{bm}
\usepackage{hyperref}
\usepackage{cleveref}
\usepackage{xcolor}

\begin{document}

\preprint{APS/123-QED}

\title{Coulomb excitation of the $\left|T_z\right|=\frac{1}{2}$, $A=23$ mirror pair}

\author{J.~Henderson}
\email{jack.henderson@surrey.ac.uk}
\affiliation{Department of Physics, University of Surrey, Guildford, GU2 7XH, United Kingdom}
\affiliation{TRIUMF, Vancouver, BC V6T 2A3, Canada}
\affiliation{Lawrence Livermore National Laboratory, Livermore, CA 94550, USA}
\author{G.~Hackman}
\affiliation{TRIUMF, Vancouver, BC V6T 2A3, Canada}
\author{P.~Ruotsalainen}
\affiliation{University of Jyv\"askyl\"a, Department of Physics, P. O. Box 35, FI-40014 University of Jyv\"askyl\"a, Finland}
\author{J.~D.~Holt}
\affiliation{TRIUMF, Vancouver, BC V6T 2A3, Canada}
\affiliation{Department of Physics, McGill University, 3600 Rue University, Montr\'eal, QC H3A 2T8, Canada}
\author{S.~R.~Stroberg}
\affiliation{TRIUMF, Vancouver, BC V6T 2A3, Canada}
\affiliation{Department of Physics, University of Washington, Seattle WA, USA}
\author{C.~Andreoiu}
\affiliation{Department of Chemistry, Simon Fraser University, Burnaby, BC V5A 1S6, Canada}
\author{G.~C.~Ball}
\affiliation{TRIUMF, Vancouver, BC V6T 2A3, Canada}
\author{N.~Bernier}
\altaffiliation{Present Address: Department of Physics \& Astronomy, University of the Western Cape, Bellville-7535, South Africa}
\affiliation{TRIUMF, Vancouver, BC V6T 2A3, Canada}
\affiliation{Department of Physics and Astronomy, University of British Columbia, Vancouver V6T 1Z1, Canada}
\author{M.~Bowry} 
\altaffiliation{Present address: University of the West of Scotland, United Kingdom}
\affiliation{TRIUMF, Vancouver, BC V6T 2A3, Canada}
\author{R.~Caballero-Folch} 
\affiliation{TRIUMF, Vancouver, BC V6T 2A3, Canada}
\author{S.~Cruz}
\affiliation{TRIUMF, Vancouver, BC V6T 2A3, Canada}
\affiliation{Department of Physics and Astronomy, University of British Columbia, Vancouver V6T 1Z1, Canada}
\author{A.~Diaz Varela}
\affiliation{Department of Physics, University of Guelph, Guelph, ON N1G 2W1, Canada}
\author{L.~J.~Evitts}
\altaffiliation{Present address: School of Computer Science and Electronic Engineering, Bangor University, Bangor, Gwynedd, LL57 2DG, UK}
\affiliation{TRIUMF, Vancouver, BC V6T 2A3, Canada}
\affiliation{Department of Physics, University of Surrey, Guildford, GU2 7XH, United Kingdom}
\author{R.~Frederick} 
\affiliation{TRIUMF, Vancouver, BC V6T 2A3, Canada}
\author{A.~B.~Garnsworthy}
\affiliation{TRIUMF, Vancouver, BC V6T 2A3, Canada}
\author{M.~Holl}
\altaffiliation{Present address: Department of Physics, Chalmers University of Technology, SE-412 96 G\"oteborg,Sweden}
\affiliation{Department of Astronomy and Physics, Saint Mary's University, Halifax, Nova Scotia B3H 3C3, Canada}
\affiliation{TRIUMF, Vancouver, BC V6T 2A3, Canada}
\author{J.~Lassen}
\affiliation{TRIUMF, Vancouver, BC V6T 2A3, Canada}
\author{J.~Measures}
\affiliation{TRIUMF, Vancouver, BC V6T 2A3, Canada}
\affiliation{Department of Physics, University of Surrey, Guildford, GU2 7XH, United Kingdom}
\author{B.~Olaizola}
\altaffiliation{ISOLDE-EP, CERN, CH-1211 Geneva 23, Switzerland}
\affiliation{TRIUMF, Vancouver, BC V6T 2A3, Canada}
\author{E.~O'Sullivan} 
\affiliation{TRIUMF, Vancouver, BC V6T 2A3, Canada}
\author{O.~Paetkau}
\affiliation{TRIUMF, Vancouver, BC V6T 2A3, Canada}
\author{J.~Park} 
\altaffiliation{Present address: Center for Exotic Nuclear Studies, Institute for Basic Science (IBS), Daejeon 34126, Republic of Korea}
\affiliation{TRIUMF, Vancouver, BC V6T 2A3, Canada}
\affiliation{Department of Physics and Astronomy, University of British Columbia, Vancouver V6T 1Z1, Canada}
\author{J.~Smallcombe}
\affiliation{TRIUMF, Vancouver, BC V6T 2A3, Canada}
\affiliation{Oliver Lodge Laboratory, University of Liverpool, Liverpool L69 7ZE, United Kingdom}
\author{C.~E.~Svensson}
\affiliation{Department of Physics, University of Guelph, Guelph, ON N1G 2W1, Canada}
\author{K.~Whitmore}
\affiliation{Department of Chemistry, Simon Fraser University, Burnaby, BC V5A 1S6, Canada}
\author{C.~Y.~Wu}
\affiliation{Lawrence Livermore National Laboratory, Livermore, CA 94550, USA}

\date{\today}

\begin{abstract}
\begin{description}
\item[Background] Electric-quadrupole ($E2$) strengths relate to the underlying quadrupole deformation of a nucleus and present a challenge for many nuclear theories. Mirror nuclei in the vicinity of the line of $N=Z$ represent a convenient laboratory for testing deficiencies in such models, making use of the isospin-symmetry of the systems.
\item[Purpose] Uncertainties associated with literature $E2$ strengths in \textsuperscript{23}Mg are some of the largest in $T_z=\left|\frac{1}{2}\right|$ nuclei in the $sd$-shell. The purpose of the present work is to improve the precision with which these values are known, to enable better comparison with theoretical models.
\item[Methods] Coulomb-excitation measurements of $^{23}$Mg and $^{23}$Na were performed at the TRIUMF-ISAC facility using the TIGRESS spectrometer. They were used to determine the $E2$ matrix elements of mixed $E2$/$M1$ transitions.
\item[Results] Reduced $E2$ transition strengths, $B(E2)$, were extracted for \textsuperscript{23}Mg and \textsuperscript{23}Na. Their precision was improved by factors of approximately six for both isotopes, while agreeing within uncertainties with previous measurements. 
\item[Conclusions] A comparison was made with both shell-model and {\it ab initio} valence-space in-medium similarity renormalization group calculations. Valence-space in-medium similarity-renormalization-group calculations were found to underpredict the absolute $E2$ strength - in agreement with previous studies.
\end{description}
\end{abstract}

\pacs{Valid PACS appear here}
\maketitle

\section{Introduction}

Electric-quadrupole ($E2$) transitions strengths are a powerful probe of nuclear structure, relating directly to the underlying quadrupole deformation of the nucleus. Simultaneously, they present a challenge to valence-space based theoretical models, with significant contributions to $E2$ strength arising from particle-hole excitations out of the model space. In the vicinity of the line of $N=Z$, mirror nuclei (nuclei with inverted numbers of protons and neutrons) are an excellent laboratory for nuclear physics, with isospin symmetry enforcing analogous structures for both nuclei. Studies of transition strengths in isobaric analogue transitions have been employed for a huge range of nuclei, from low-mass systems such as \textsuperscript{7}Be and \textsuperscript{7}Li~\cite{ref:SHenderson_19}, through the $f_{7/2}$ shell (e.g. Ref.~\cite{ref:Boso_19}), and extending into the upper-$fp$ and $g_{9/2}$ model spaces (e.g. Ref.~\cite{ref:Morse_18}). 

Within the $sd$-shell, one is able to compare modern {\it ab initio} techniques such as the valence-space in-medium similarity renormalization group (VS-IMSRG) to calculations utilising exceptionally successful empirical shell-model interactions such as the USDB~\cite{ref:USDB}. Systematic studies of deficiencies in such models require, however, high-quality experimental data. In this work, we build on our previous studies of \textsuperscript{22}Mg~\cite{ref:Henderson_18} and \textsuperscript{21}Mg~\cite{ref:Ruotsalainen_19} by presenting an improved experimental measurement of the low-lying $E2$ strength in the $\left|T_z\right|=\frac{1}{2}$, $A=23$ mirror pair, \textsuperscript{23}Mg and \textsuperscript{23}Na. Prior to the present work, the $B(E2)$ value between the ground and first-excited state in \textsuperscript{23}Mg~\cite{ref:Warburton_74,ref:Tikkanen_90} was the most imprecisely measured of all $T_z=-\frac{1}{2}$, $sd$-shell nuclei~\cite{ref:ENSDF}. A detailed systematic study, comparing VS-IMSRG and shell-model calculations to the available data within the $sd$-shell is the subject of a separate publication~\cite{ref:Henderson_21}.

The precision to which $E2$ strengths are determined in odd-mass $sd$-shell nuclei is often limited by the fact that decays are of a mixed $E2/M1$ nature. When the decay is dominated by $M1$ strength, as is the case in $^{23}$Mg and $^{23}$Na, the leading uncertainty in determining the $E2$ strength is typically the mixing ratio $\delta$ between $E2$ and $M1$ contributions determined, for example, from the angular correlations between emitted $\gamma$ rays. By performing a Coulomb excitation measurement, rather than determining the $E2$ strength from the decay properties, this source of uncertainty can be largely eliminated, allowing for a higher level of precision. 

\section{Experimental Details}

\begin{figure}
\centerline{\includegraphics[width=.95\linewidth]{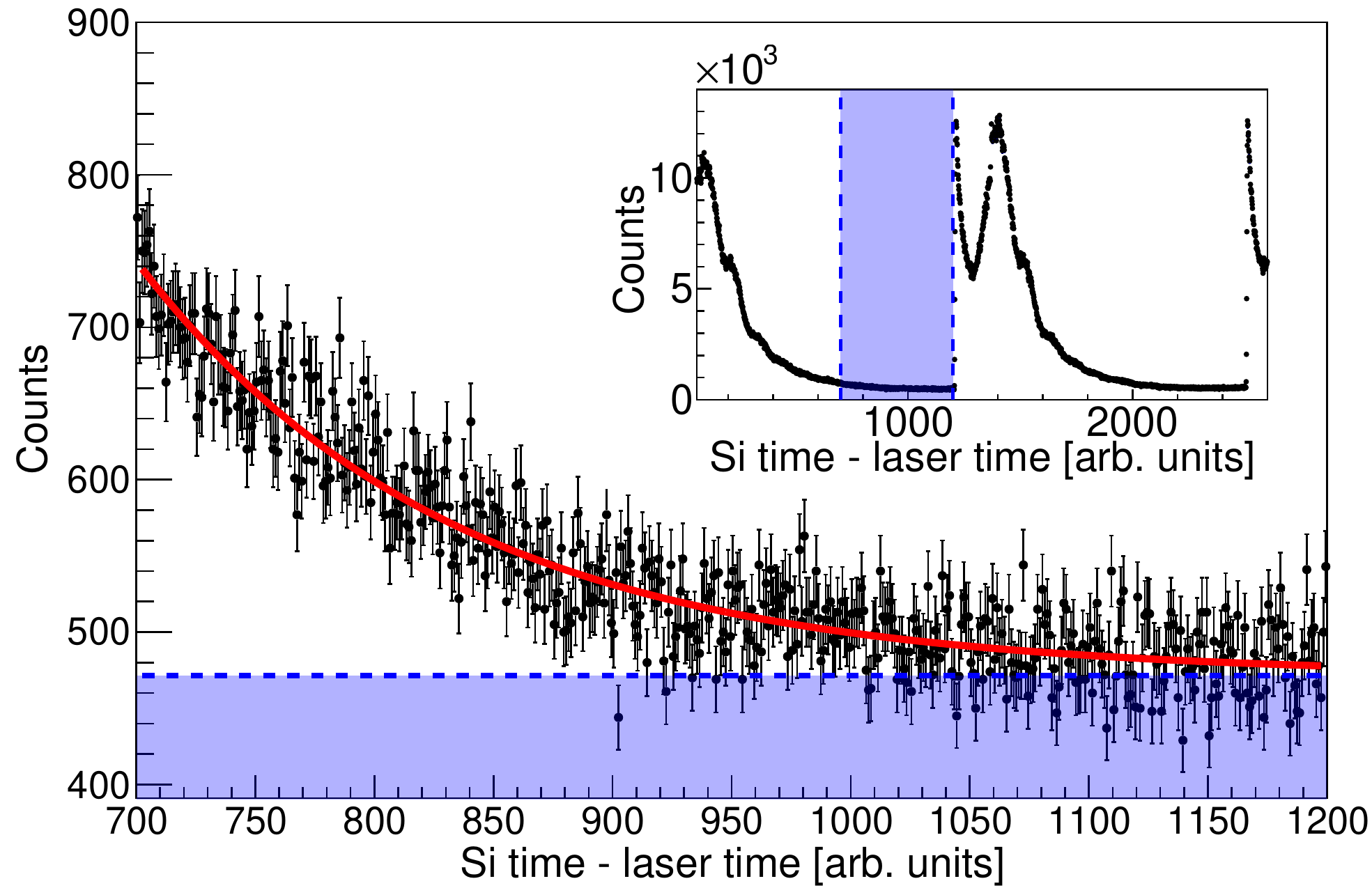}}
\caption{Tail of the silicon-time laser-time distribution (see text for details), least-squares fit with an exponential plus constant background. The integral of the background is used to determine the surface ionized contamination originating from $^{23}$Na. Shown in the inset is the total time structure arising from the laser-ionization in the present measurement with the fitted area indicated.}
\label{fig:Lasers}
\end{figure}

$^{23}$Mg and $^{23}$Na were investigated through Coulomb excitation using the TIGRESS facility~\cite{ref:Hackman_14} at TRIUMF ISAC. $^{23}$Mg nuclei were produced by the impinging of 480-MeV protons onto a SiC ISAC target. The Mg atoms produced were then selectively laser ionized using three step resonant excitation (285.3~nm-880.8~nm-291.6~nm) into an auto-ionizing state and extracted. $^{23}$Na contamination was suppressed by the use of the ion-guide laser ion source (IG-LIS)~\cite{ref:Raeder_14}. A repeller plate is held at 40~V to suppress the extraction of surface-ionized contaminants by factors of up to $10^6$. $^{23}$Na ions were produced by the surface ion source of the TRIUMF offline ion source (OLIS)~\cite{ref:Jayamanna_08}. The beams were then accelerated by the TRIUMF ISAC accelerator chain and delivered to TIGRESS. The $^{23}$Mg/$^{23}$Na cocktail beam had an energy of 42.9~MeV, while the $^{23}$Na beam provided by OLIS was provided at energies of both 42.9~MeV and 39.4~MeV. The total beam intensity for the $^{23}$Mg portion of the experiment was maintained at roughly $3\cdot10^5$~particles per second - this includes a component from the remaining $^{23}$Na contamination. The $^{23}$Na beam intensity was maintained at approximately $6\cdot10^7$~particles per second. The beams were then impinged onto a 0.44-mg/cm$^2$ thick, $^\text{nat}$Ti target at the center of the TIGRESS array. Scattered beam- and target-like nuclei were detected in an S3-type~\cite{ref:Micron} silicon detector, mounted 31-mm downstream of the target position. Gamma rays were detected using the TIGRESS array, which for the present measurement comprised fourteen clover-type HPGe detectors. The HPGe detectors were operated in their withdrawn configuration, with the face of the detectors 14.5~cm from the target and the BGO suppression shields forward, providing the best possible peak-to-background ratio and Doppler-correction.

\begin{figure}
\centerline{\includegraphics[width=\linewidth]{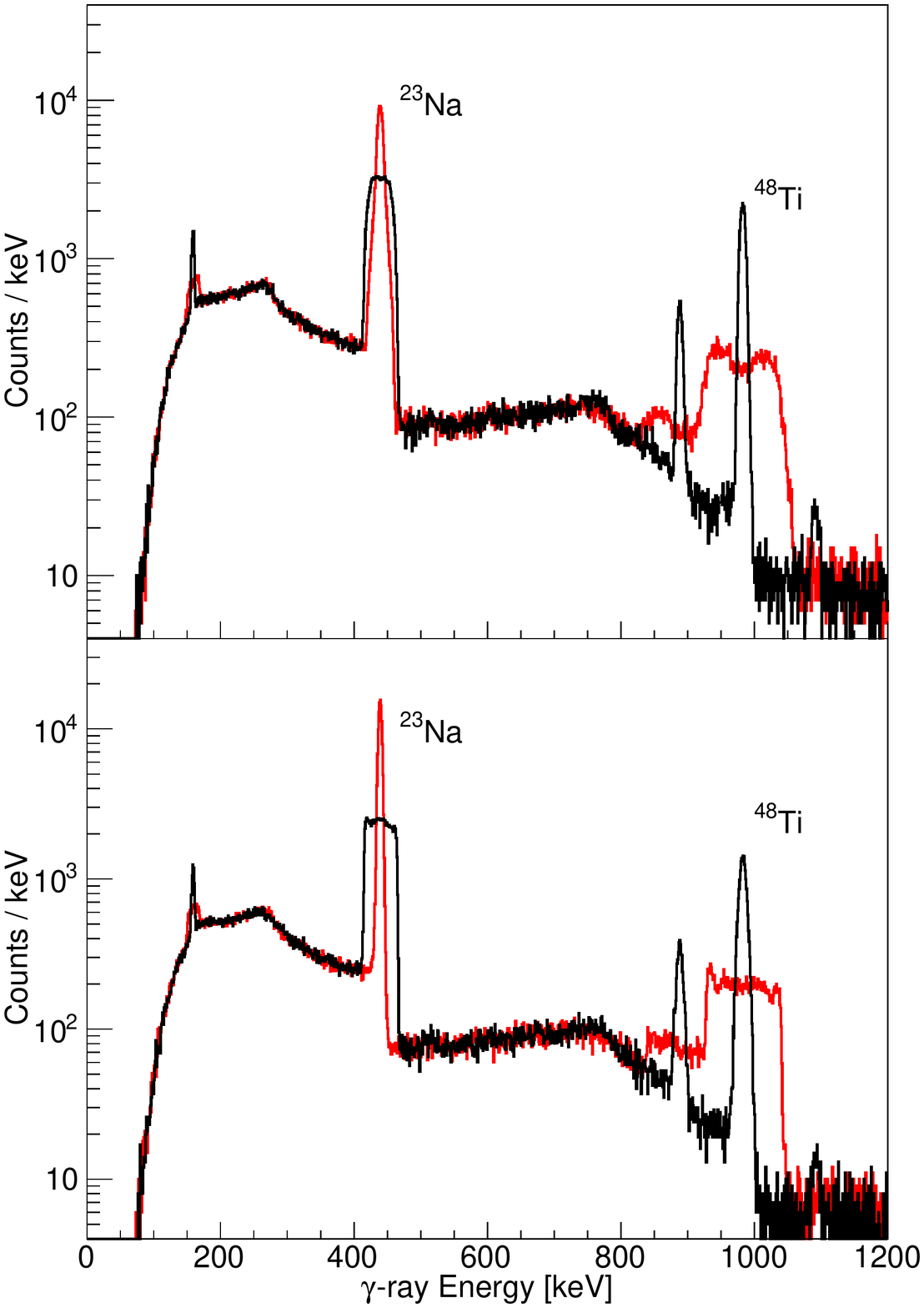}}
\caption{Doppler-corrected $\gamma$-ray spectra on the basis of $^{23}$Na (red) and $^{48}$Ti (black) kinematics for a $^{23}$Na beam energy of 39.4~MeV. Top: Detection of a target-like recoil ($^{48}$Ti) in the downstream annular silicon detector. Bottom: Detection of a beam-like recoil ($^{23}$Na) in the downstream annular silicon detector. The additional width of the $^{23}$Na peak in the top figure arises from the wide angles at which the scattering occurs, leading to significant slowing in the target material. Other lines in the titanium corrected (black) spectra arise from isotopes of titanium with a lower natural abundance than $^{48}$Ti (73.8\%). }
\label{fig:Na39MeV_Comb}
\end{figure}

\begin{figure}
\centerline{\includegraphics[width=\linewidth]{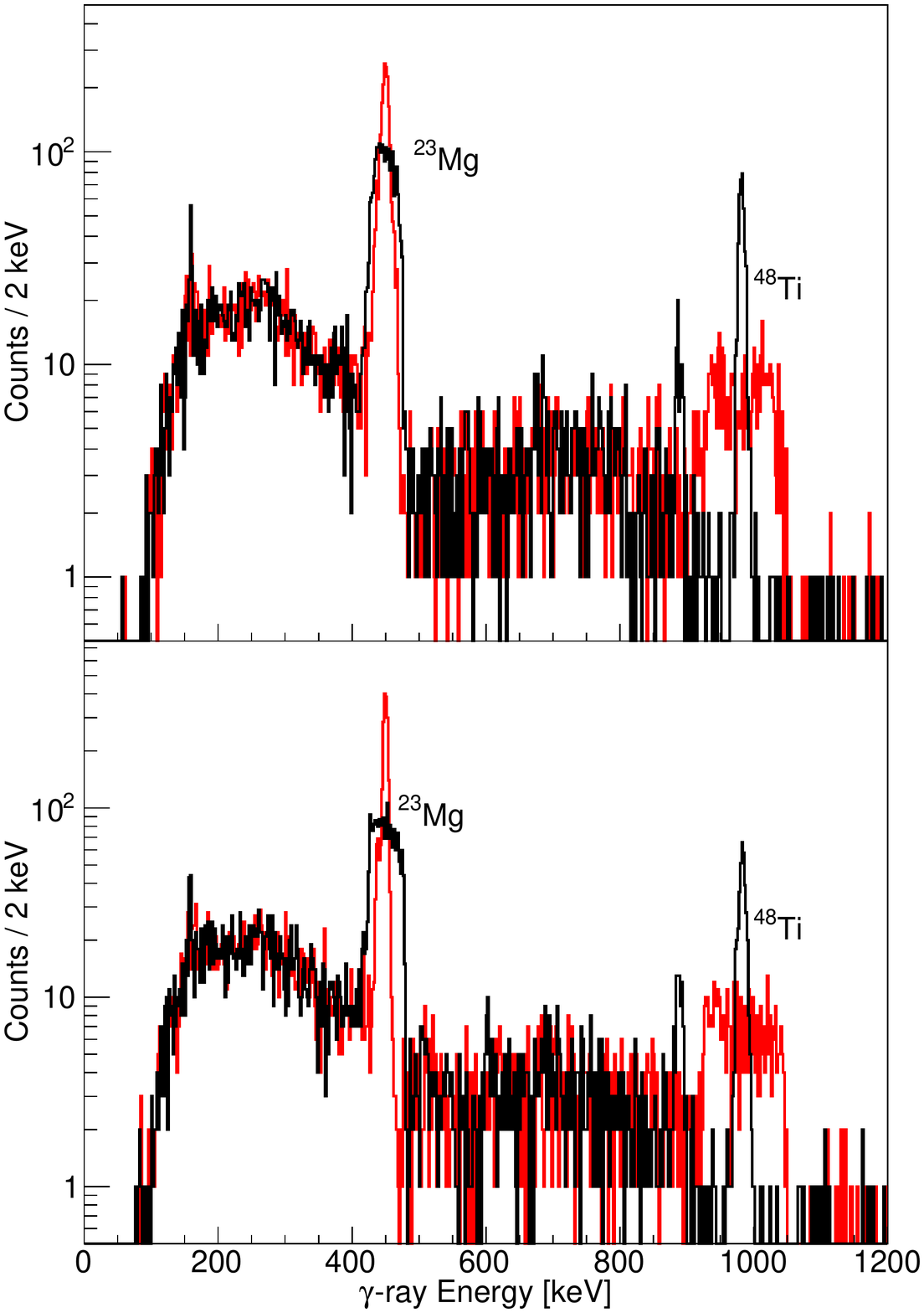}}
\caption{As Fig.~\ref{fig:Na39MeV_Comb} but for a cocktail $^{23}$Mg ($\approx85\%$) and $^{23}$Na ($\approx15\%$) beam at an energy of 42.9~MeV. }
\label{fig:MgNa_Comb}
\end{figure}

While the use of IG-LIS heavily suppresses extraction of $^{23}$Na, a degree of contamination remains which was monitored in two ways. First, a Bragg detector 
was used to provide an instantaneous measure of the beam composition. While the composition is being determined in this way experimental data cannot be acquired. For the second method, the 10 kHz signal used to synchronize the laser ionization system was used, with every second pulse triggering the generation of a ramping waveform, which could then be digitized. The amplitude of the digitized waveform thereby gave a proxy for the time of the detection relative to the laser-ionization pulse and could thus be used to distinguish laser-ionized beam components which had a 10~kHz pulsed structure from the continuous surface ionized contaminants. This method allowed for a continuous determination of contamination, allowing to monitor for sudden changes in the ISAC target behavior. Based on these analyses, the $^{23}$Na contribution to the beam cocktail was determined to be 15.2(9)~\% of the total, with the uncertainty being predominantly systematic and arising from the choice of fitting region. Figure~\ref{fig:Lasers} shows the laser timing distribution, the tail of which was fit with an exponential and baseline to determine the relative contributions to the beam cocktail. 

\section{Analysis}

\begin{figure}
\centerline{\includegraphics[width=\linewidth]{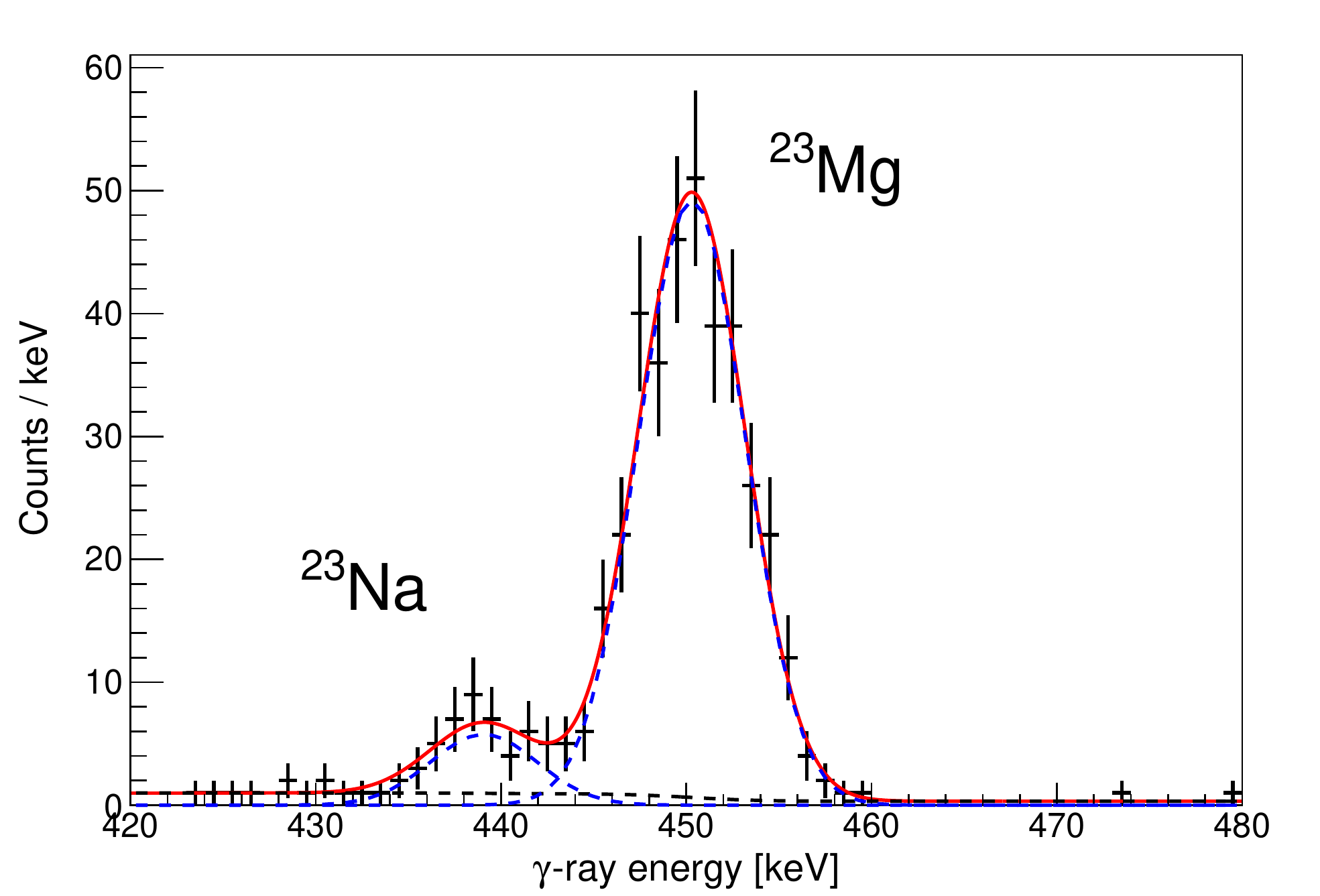}}
\caption{Fit of the $\gamma$-ray peaks observed in TIGRESS corresponding to the de-excitation of the first-excited state in $^{23}$Mg and the analogue state in the stable contaminant and mirror nucleus, $^{23}$Na. These data were coincident with events from the first four rings of the downstream annular silicon detector, corresponding to angles of $19.5^\circ\rightarrow25.8^\circ$. This fitting method can be used for all cases where the beam-like particle was detected. See the text for details of the analysis for target-like particle detection.}
\label{fig:MgNa_Fit}
\end{figure}

The data were unpacked using the {\small GRSISort}~\cite{ref:GRSISort} software package, built in a {\small ROOT}~\cite{ref:ROOT} framework. Gamma-ray events were Doppler corrected event-by-event on the basis of the beam and target kinematics determined from the hit location in the annular silicon detectors and whether the detected particle had beam-like or target-like kinematic properties. Gamma-ray spectra for \textsuperscript{23}Na at 39.4~MeV, and the \textsuperscript{23}Mg + \textsuperscript{23}Na cocktail beam are shown in Fig.~\ref{fig:Na39MeV_Comb} and Fig.~\ref{fig:MgNa_Comb}, respectively. Relative $\gamma$-ray detection efficiencies for TIGRESS were determined using a standard suite of $^{152}$Eu, $^{133}$Ba and $^{60}$Co sources. $^{23}$Na data were split into forty-eight groups: twelve angular bins for both beam-like and target-like detection, repeated for both beam energies. The $^{23}$Mg data were binned in twelve groups, six angular groups each for beam-like and target-like scattering. Yields were adjusted for the natural abundance of $^{48}$Ti. 

In the beam-like scattering data the $^{23}$Mg and $^{23}$Na $\gamma$-ray lines were readily distinguished and were fitted individually, as shown in Fig.~\ref{fig:MgNa_Fit}. The observed $^{48}$Ti yield was then adjusted for the observed $^{23}$Na component on the basis of the 42.9~MeV $^{23}$Na data. For the target scattering data the two components of the $A=23$ $\gamma$-ray peak were not always distinguishable. The $^{23}$Na component was therefore determined and subtracted on the basis of the observed component in the beam-like scattering data and of the 42.9~MeV, pure $^{23}$Na data taken with OLIS. $^{23}$Na contamination could thereby be handled empirically, without requiring assumptions about beam composition and minimizing the introduction of systematic uncertainties.

\begin{figure}
\centerline{\includegraphics[width=\linewidth]{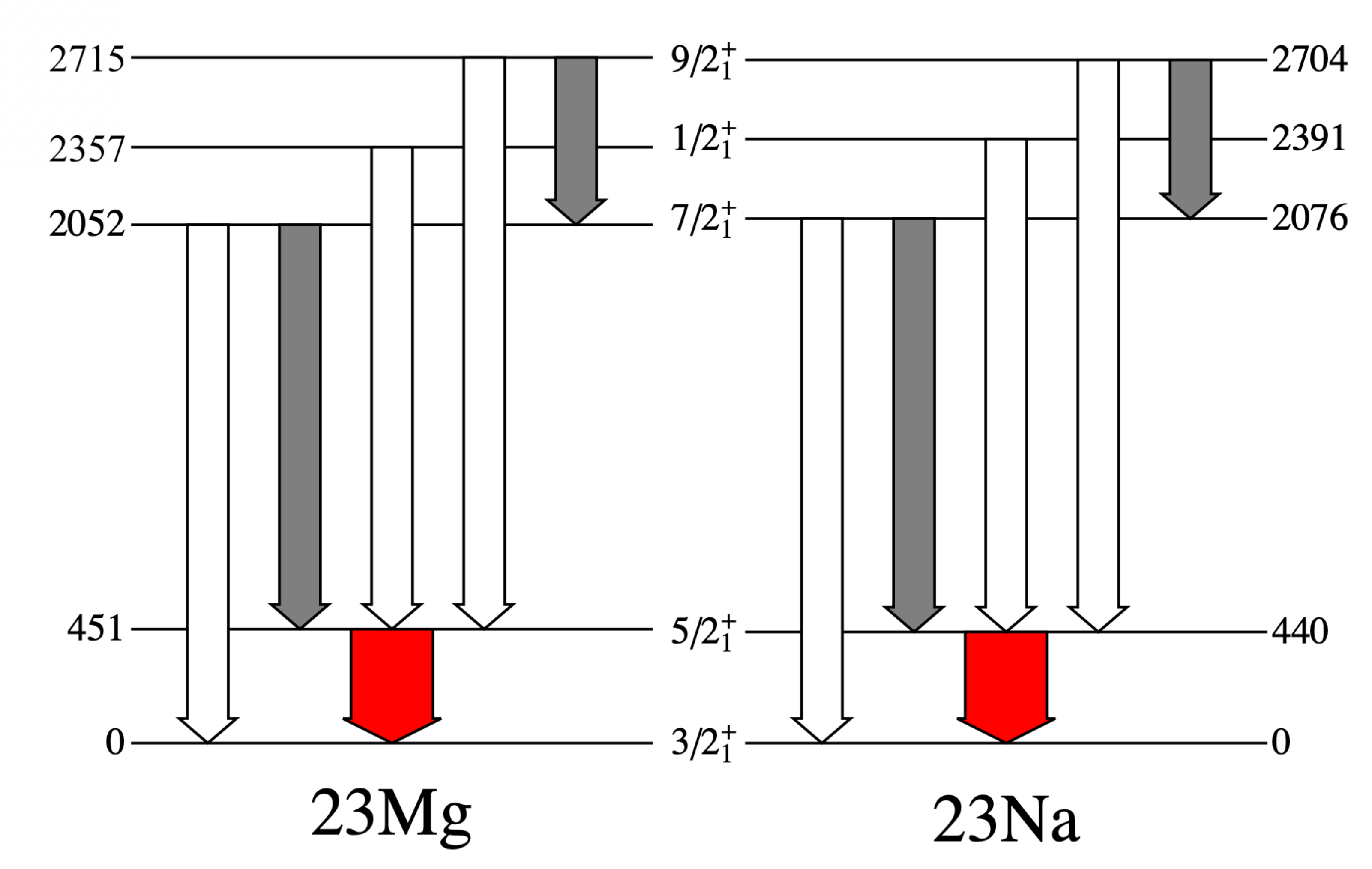}}
\caption{Low-lying levels in $^{23}$Mg and $^{23}$Na relevant to the present analysis. The $5/2^{+}\rightarrow3/2^{+}$ transition (red) was investigated and other transitions were included within the GOSIA analysis. Gray transitions indicate mixed $E2/M1$. Data taken from Ref.~\cite{ref:ENSDF}.}
\label{fig:Levels}
\end{figure}

\begin{figure}
\centerline{\includegraphics[width=.98\linewidth]{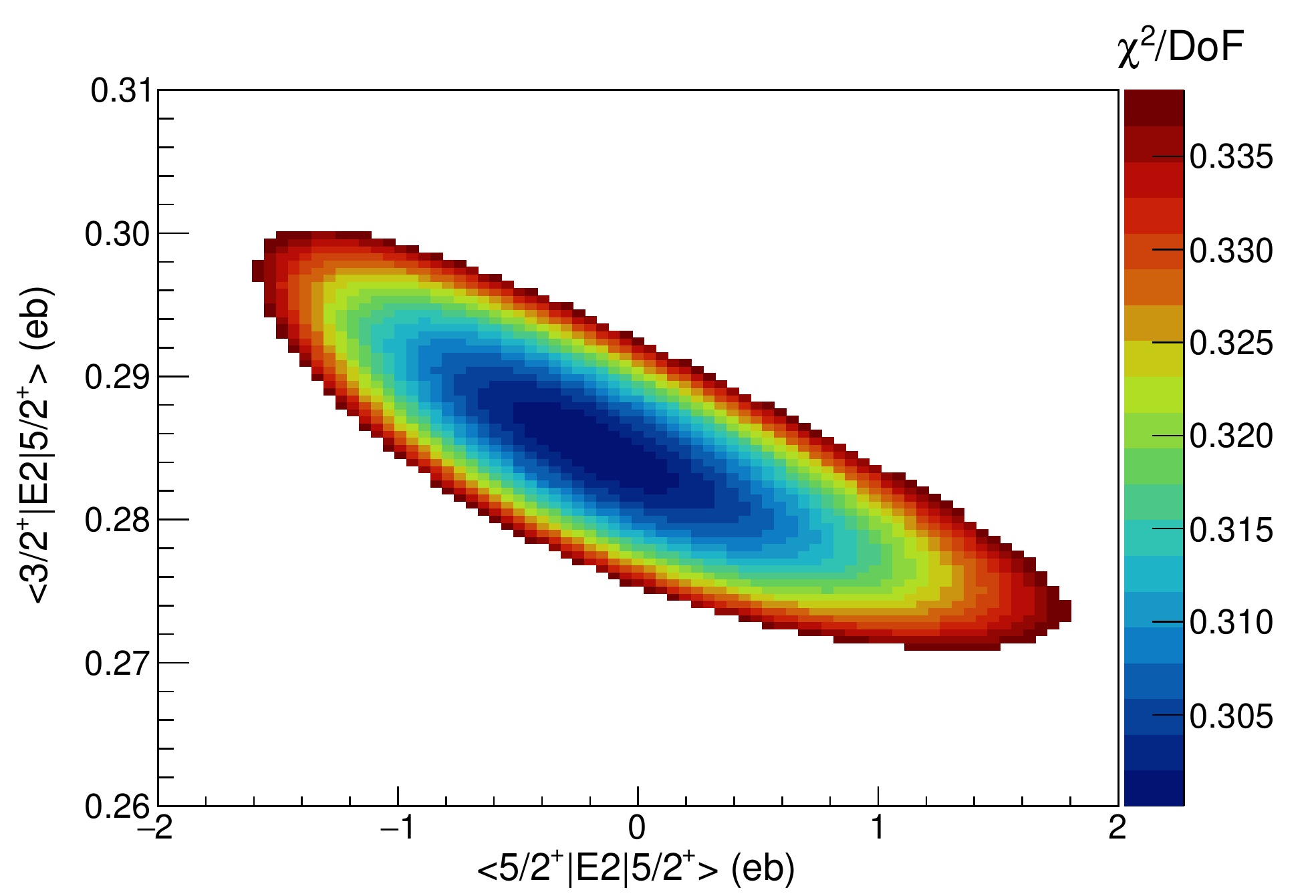}}
\caption{$\chi^2$ surface resulting from the GOSIA2 analysis of $^{23}$Mg from which transition and diagonal matrix-elements were extracted.}
\label{fig:Mg23ChisqSurf}
\end{figure}

\begin{figure}
\centerline{\includegraphics[width=\linewidth]{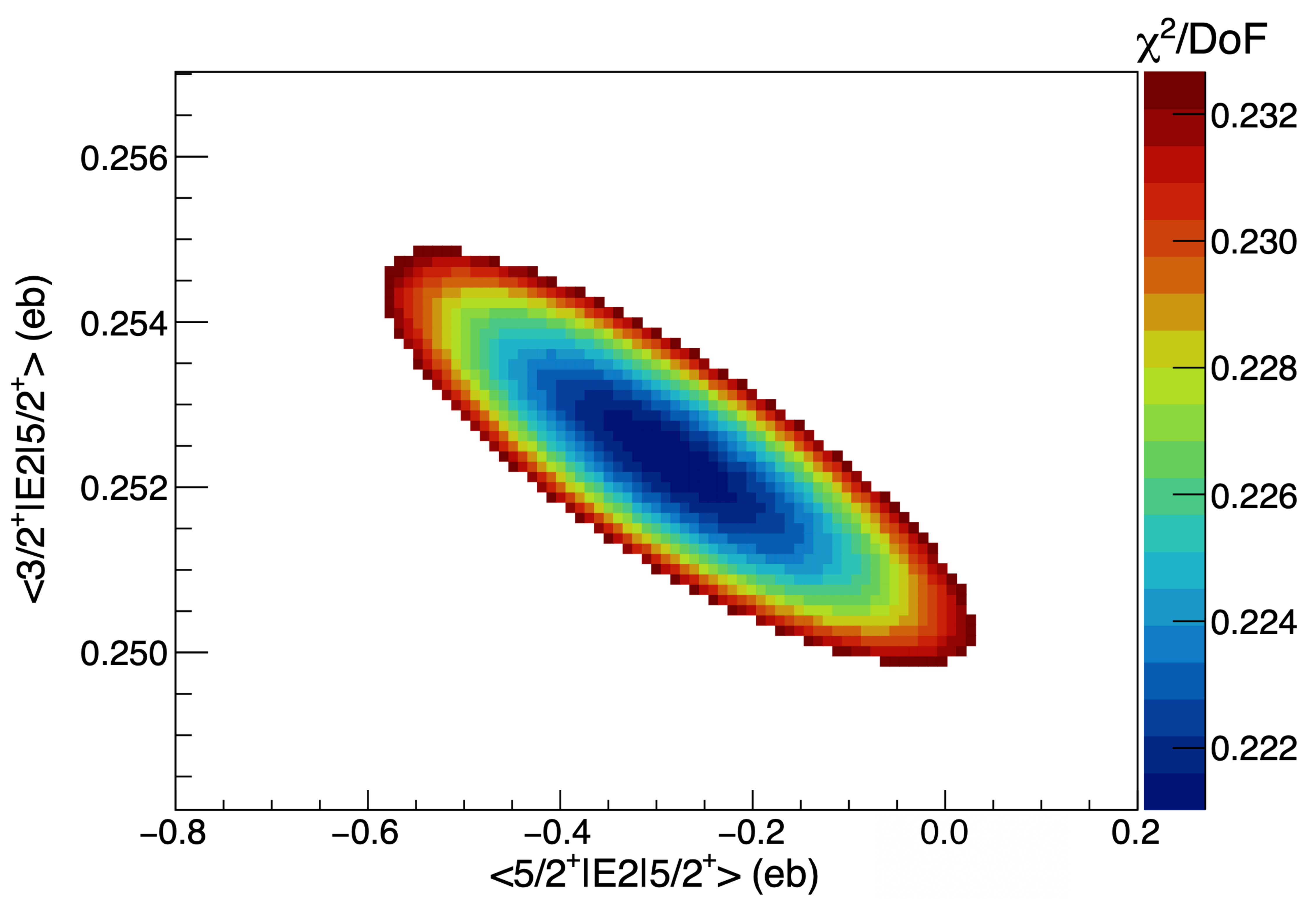}}
\caption{$\chi^2$ surface resulting from the GOSIA2 analysis of $^{23}$Na from which transition and diagonal matrix-elements were extracted.}
\label{fig:Na23ChisqSurf}
\end{figure}

The Coulomb-excitation analysis was performed in the coupled-channels GOSIA2 code~\cite{ref:GOSIA} used to simultaneously analyze beam- and target-like data. The levels included in the GOSIA2 analysis are shown in Fig.~\ref{fig:Levels}. Ground-state spectroscopic quadrupole moments for both $^{23}$Na and $^{23}$Mg were taken at their evaluated values~\cite{ref:Stone_16}. For each beam all data were analyzed simultaneously, maximizing sensitivity. The $\bra{\frac{3}{2}^+}E2\ket{\frac{5}{2}^+}$ and $\bra{\frac{5}{2}^+}E2\ket{\frac{5}{2}^+}$ matrix elements were varied in order to construct $\chi^2$ surfaces to incorporate any mutual dependence. $\chi^2$ surfaces for $^{23}$Mg and $^{23}$Na are shown in Fig.~\ref{fig:Mg23ChisqSurf} and Fig.~\ref{fig:Na23ChisqSurf}, respectively. Little sensitivity was found to the diagonal matrix element beyond an indication of the sign in \textsuperscript{23}Na. Matrix elements to higher-lying states were fixed to their literature values during the minimization procedure, however their $1\sigma$ limits were investigated to quantify any impact on the result and are incorporated as a systematic uncertainty. 

\section{Discussion}

Extracted matrix elements are summarized in Tab.~\ref{tab:MatrixElements}, along with other properties derived from the present results. We compare the present results with those calculated from two theoretical models. VS-IMSRG calculations were performed using the EM1.8/2.0 interaction~\cite{ref:Hebeler_11,ref:Simonis_16}, which was generated by SRG evolution~\cite{ref:Bogner_07} of the chiral N$^3$LO NN interaction of Entem and Machleidt~\cite{ref:Entem_03}, and adding a non-locally regulated N$^2$LO 3N interaction with the low energy constants adjusted to reproduce the triton binding energy and the $^{4}$He matter radius. Calculations are performed in a harmonic oscillator basis of $\hbar\omega=20$~MeV with $2n+\ell\leq e_{max}$=12
and with a truncation on the three body matrix elements $e_1+e_2+e_3\leq E_{3max}$=16. All operators are truncated at the normal-ordered two-body level. A diagonalization was then performed using the NuShellX~\cite{ref:NushellX} code.  Shell-model calculations were also performed in NuShellX, making use of the USDB interaction~\cite{ref:USDB} with effective charges of $e_\pi=1.36$ and $e_\nu=0.45$.
\begin{table*}
\caption{$E2$ matrix elements, $B(E2)$ values, spectroscopic quadrupole moments and mixing ratios deduced from the present work with statistical and systematic uncertainties quoted, in that order. Where available, literature values are shown for comparison. Mixing ratios were deduced one the basis of the literature lifetimes and the presently determined $B(E2)$ values.}
\label{tab:MatrixElements}
\begin{ruledtabular}
\begin{tabular}{llll}
$^{23}$Na 										& This Work 								& Literature 					& Ref. \\
\hline \\[-6pt]
$\bra{\frac{3}{2}^+_1}E2\ket{\frac{5}{2}^+_1}$ eb 			& $0.252\pm0.003\pm0.004$ 					& $0.237^{+0.014}_{-0.015}$ 		& \cite{ref:ENSDF}\\[2pt] 
$B(E2;\frac{5}{2}^+\rightarrow\frac{3}{2}^+)$ e$^2$fm$^4$ 	& $106\pm3\pm3$ 							& $93\pm12$ 					& \cite{ref:ENSDF} \\[2pt]
$\bra{\frac{5}{2}^+_1}E2\ket{\frac{5}{2}^+_1}$ eb 			& $-0.29^{+0.32}_{-0.29}\pm0.05$ 				& 							& \\[2pt]
$Q_s(\frac{5}{2}^+_1)$ eb 							& $-0.22^{+0.25}_{-0.22}\pm0.04$ 				& 							& \\[2pt]
$\delta^2_{E2/M1}$ 									& $0.0038\pm0.0004$ 						& $0.0034^{+0.0004}_{-0.0003}$ 	& \cite{ref:ENSDF} \\[2pt]
\hline \\[-7pt]
$^{23}$Mg & & & \\
\hline \\[-6pt]
$\bra{\frac{3}{2}^+_1}E2\ket{\frac{5}{2}^+_1}$ eb 			& $0.285\pm0.015\pm0.004$ 					& $0.23^{+0.07}_{-0.10}$ 			& \cite{ref:ENSDF}\\[2pt] 
$B(E2;\frac{5}{2}^+\rightarrow\frac{3}{2}^+)$ e$^2$fm$^4$ 	& $135^{+15}_{-14}\pm4$ 					& $86\pm58$ 					& \cite{ref:ENSDF} \\[2pt]
$\bra{\frac{5}{2}^+_1}E2\ket{\frac{5}{2}^+_1}$ eb 			& $-0.2^{+2.0}_{-1.3}\pm0.05$ 					& 							& \\[2pt]
$Q_s(\frac{5}{2}^+_1)$ eb 							& $-0.15^{+1.50}_{-1.00}\pm0.04$ 				& 							& \\[2pt]
$\delta^2_{E2/M1}$ 									& $0.0056\pm0.0006$						& $0.0036^{+0.0028}_{-0.0020}$	& \cite{ref:ENSDF} \\[2pt]
\end{tabular}
\end{ruledtabular}
\vspace{-10pt}
\end{table*}

\begin{table}
\caption{$B(E2$ values determined in the present work compared to those calculated using the VS-IMSRG method and the nuclear shell model using the USDB interaction.}
\label{tab:Comparison}
\begin{ruledtabular}
\begin{tabular}{lcccccc}
 & & &  \multicolumn{2}{c}{B(E2)$\downarrow$ [e$^2$fm$^4$]} & \\
 \hline \\[-7pt]
Isotope & $J^\pi_i$ & $J^\pi_f$ & Expt. & VS-IMSRG & USDB & \\[+2pt]
 \hline \\[-7pt]
$^{23}$Mg 	& $\frac{5}{2}^+_1$ 	& $\frac{3}{2}^+_1$ 	& 135 (15) 	& 75.2 	& 117.3		\\[+1pt]
$^{23}$Na		& $\frac{5}{2}^+_1$ 	& $\frac{3}{2}^+_1$ 	& 106 (4) 	& 56.9 	& 109.1 		\\[+1pt]
\end{tabular}
\end{ruledtabular}
\end{table}

Table~\ref{tab:Comparison} shows the present results compared to those calculated using the aforementioned models. The shell-model (USDB) calculations well reproduce the observed $B(E2)$ values. VS-IMSRG values, meanwhile, are considerably lower than the experimentally determined ones. This deficiency is consistent with that observed in our previous studies of $\left|T_z\right|=1$ mirror pairs~\cite{ref:Henderson_18}. While the VS-IMSRG values are deficient, it should be noted that the relative $B(E2)$ strengths are better reproduced by the {\it ab initio} calculations. Defining the ratio $R=\frac{B(E2\frac{5}{2}^+\rightarrow\frac{3}{2}^+)[^{23}\text{Mg}]}{B(E2\frac{5}{2}^+\rightarrow\frac{3}{2}^+)[^{23}\text{Na}]}$, we find that $R_{exp}=1.27(14)$, whereas $R_{USDB}=1.06$ and $R_{VS-IMSRG}=1.34$. In order to understand the relative behaviour of $E2$ strengths across mirror-pairs, a systematic study is required, which is the subject of a separate work~\cite{ref:Henderson_21}.

\section{Conclusions}

\textsuperscript{23}Mg and \textsuperscript{23}Na have been studied by Coulomb excitation using particle-$\gamma$ coincidences at TRIUMF-ISAC. The relative insensitivity of the Coulomb excitation methodology to the $M1$ transitions which dominate the decay of the first excited states allowed for the extraction of $E2$ transition strengths with superior precision to that previously achieved, while agreeing within uncertainties with literature values. Calculations were performed, employing both the shell-model with the USDB interaction, and the {\it ab initio} VS-IMSRG methodology. Consistent with previous work, it was found that the VS-IMSRG calculations significantly underpredict the $E2$ transition strength. A detailed, systematic investigation of deficiencies in $E2$ strength from VS-IMSRG calculations is the subject of a separate study~\cite{ref:Henderson_21}.

\section{Acknowledgements}

The authors would like to thank the TRIUMF beam delivery group for their efforts in providing high-quality stable and radioactive beams. This work has been supported by the Natural Sciences and Engineering Research Council of Canada (NSERC), The Canada Foundation for Innovation and the British Columbia Knowledge Development Fund. TRIUMF receives federal funding via a contribution agreement through the National Research Council of Canada. Computations were performed with an allocation of computing resources on Cedar at WestGrid and Compute Canada, and on the Oak Cluster at TRIUMF managed by the University of British Columbia department of Advanced Research Computing (ARC). Work at LLNL was performed under contract DE-AC52-07NA27344. This work was supported by the Office of Nuclear Physics, U.S. Department of Energy, under grants desc0018223 (NUCLEI SciDAC-4 collaboration) and by the Field Work Proposal ERKBP72 at Oak Ridge National Laboratory (ORNL).
SRS is supported by the U.S. Department of Energy under contract DE-FG02-97ER41014. JH is supported at the University of Surrey under UKRI Future Leaders Fellowship grant no. MR/T022264/1.

Fig.~\ref{fig:Levels} was created using the SciDraw scientific figure preparation system~\cite{ref:SciDraw}.
The codes imsrg++~\cite{ref:imsrgcode} and nutbar~\cite{ref:nutbar} used in this work make use of the Armadillo library~\cite{ref:Armadillo}.

\bibliography{mg23}

\end{document}